\documentclass[aps,prd,twocolumn,superscriptaddress,nofootinbib,showpacs,floatfix]{revtex4-2}

\usepackage[utf8]{inputenc}
\usepackage[T1]{fontenc}
\usepackage{lmodern}
\usepackage{amsmath,amssymb}
\usepackage{graphicx}
\DeclareGraphicsExtensions{.pdf,.png}
\usepackage{booktabs}
\usepackage{multirow}
\usepackage{xcolor}
\usepackage[colorlinks=true,linkcolor=blue!60!black,citecolor=blue!60!black,urlcolor=blue!60!black]{hyperref}
\UseRawInputEncoding

\newcommand{\LCDM}{\ensuremath{\Lambda\mathrm{CDM}}}
\newcommand{\dchi}{\ensuremath{\Delta\chi^{2}}}
\newcommand{\kms}{\ensuremath{\mathrm{km\,s^{-1}\,Mpc^{-1}}}}
\newcommand{\HM}[1]{H_{\rm M}^{2}(#1)}
\newcommand{\Aeff}{\ensuremath{A_{\rm eff}}}

\newcommand{\figorbox}[1]{%
  \IfFileExists{#1.pdf}{\includegraphics[width=\columnwidth]{#1.pdf}}{%
  \IfFileExists{#1.png}{\includegraphics[width=\columnwidth]{#1.png}}{%
  \fbox{\parbox[c][0.55\columnwidth][c]{0.95\columnwidth}{\centering
  \small Figure file \texttt{\detokenize{#1}.pdf} not found.\\ Upload it to the project.}}}}}

\begin{document}

\title{Non-Markovian Cosmology with Causal Memory Kernels}

\author{Aritra Sanyal}
\email{aritrasanyal1@gmail.com}
\affiliation{Department of Mathematics, Jadavpur University, Kolkata-700 032, India}

\author{Prabir Rudra}
\email{prudra.math@gmail.com}
\affiliation{Department of Mathematics, Asutosh College, Kolkata-700 026, India}

\date{September 30, 2026}

\begin{abstract}
We investigate a non-Markovian extension of Friedmann cosmology in which the
present expansion rate retains information about the past cosmic evolution
through a causal memory kernel, with \LCDM{} recovered in the limit of vanishing
kernel amplitude, $A\to0$. We show that five kernels (Linear, Power Law,
Exponential, Gaussian and Oscillatory) arise as cosine transforms of the
spectral densities of an unresolved hidden sector, and that the Exponential and
Linear kernels also follow independently from Israel-Stewart causal viscosity
and from Deser-Woodard non-local gravity, respectively. We constrain the
kernels with 30 Cosmic Chronometer (CC) measurements and with $H(z)$ derived
from the DESI DR1 and DR2 baryon acoustic oscillation (BAO) data, using
multi-start $\chi^{2}$ minimisation followed by Markov chain Monte Carlo
sampling, and compare them with \LCDM{} through the Akaike and Bayesian
information criteria. The Linear amplitude is consistent with zero
($A=0.002\pm0.030$ for CC+DR2). The Power-law, Exponential and Gaussian kernels
are observationally degenerate: the data select sharply peaked kernels that
constrain only an integrated amplitude $\Aeff\simeq0.01$--$0.03$, equivalent to
a uniform $0.5$--$1.5\%$ rescaling of $H(z)$, and they are disfavoured relative
to \LCDM{} ($\Delta{\rm AIC}=+2.5$ to $+3.6$). Only the Oscillatory kernel
improves the fit appreciably ($\dchi=4.5$--$8.1$), but it requires a negative
damping rate, which is incompatible with the causal derivations, and it is not
favoured once CC data are included ($\Delta{\rm BIC}=+4.4$ to $+6.2$).
\end{abstract}

\pacs{98.80.-k, 95.36.+x, 98.80.Es, 04.50.Kd}

\maketitle

\section{Introduction}
\label{sec:intro}

The standard cosmological model, built on the Friedmann--Lema\^{i}tre--Robertson--Walker
(FLRW) spacetime, is inherently Markovian: the Hubble parameter $H(z)$ depends
only on the instantaneous energy density, with no dependence on the past
evolution of the Universe. This is an assumption rather than a consequence of
general relativity. The FLRW description is a coarse-grained one, in which
short-wavelength fluctuations, sub-horizon quantum modes and any unresolved
components have been averaged out, and eliminating degrees of freedom coupled to
a macroscopic variable generically produces a memory term in its equation of
motion~\cite{Zwanzig1961,Mori1965,Caldeira1983a}. A memoryless description is
recovered only when the correlation time of the eliminated environment is short
compared with the Hubble time. Memory effects are further suggested by the
non-local effective actions generated by quantum loop
corrections~\cite{Deser2007}, by causal dissipative fluids~\cite{Israel1976,IsraelStewart1979},
and by the dependence of the averaged expansion on the history of structure
formation~\cite{Weinberg2008,Capozziello2011}. In such scenarios the present
expansion rate can retain information about past cosmic evolution, leading
naturally to non-Markovian dynamics.

On the observational side, the Hubble and $S_8$ tensions, together with the
hints of evolving dark energy in recent DESI data~\cite{DESI2024,DESI2025},
motivate asking whether the instantaneous relation between $H(z)$ and the energy
density is overly rigid. A non-Markovian extension contains \LCDM{} as a
limiting case, which allows a nested model comparison. It can generate
structure in $H(z)$ and effective phantom-like behaviour without introducing new
fields, and it makes history-dependent predictions that instantaneous models
cannot.

Motivated by these considerations, we develop a non-Markovian extension of the
Friedmann equation governed by a causal memory kernel. We study five physically
motivated continuous kernels, derive their microscopic origin, and constrain
their parameters with Cosmic Chronometer (CC) measurements and DESI BAO data
from Data Release~1 (DR1, 2024) and Data Release~2 (DR2, 2025). Beyond
best-fit values, we obtain posterior distributions by Markov chain Monte Carlo
(MCMC) sampling and assess whether the additional kernel parameters are
warranted using information criteria.

\section{The Markovian Friedmann equation}
\label{sec:markov}

For a spatially flat FLRW universe the first Friedmann equation reads
\begin{equation}
  H^{2}(z) = \HM{z} \equiv H_0^{2}\left[\Omega_m(1+z)^{3}+\Omega_\Lambda\right],
  \label{eq:friedmann}
\end{equation}
where $H_{\rm M}(z)$ denotes the Markovian (\LCDM) expansion rate, $H_0$ is the
present-day Hubble constant and $\Omega_m$ and $\Omega_\Lambda$ are the
dimensionless matter and dark-energy density parameters. This formulation is
strictly Markovian: $H(z)$ depends only on the instantaneous energy content of
the Universe. Throughout, we adopt the Planck 2018 fiducial
values~\cite{Planck2018} $H_0=67.4~\kms$, $\Omega_m=0.315$ and
$\Omega_\Lambda=0.685$.

\section{Non-Markovian extension}
\label{sec:nonmarkov}

We generalise Eq.~\eqref{eq:friedmann} by adding a causal memory integral,
\begin{equation}
  H^{2}(z) = \HM{z} + \int_{0}^{z} K(z-z')\,\HM{z'}\,dz'.
  \label{eq:nm_friedmann}
\end{equation}
Causality requires $K(z-z')=0$ for $z'>z$. The kernel $K(\Delta z)$, with
$\Delta z\equiv z-z'$, encodes the strength and redshift scale of the cosmic
memory, and standard \LCDM{} is recovered for $K\to0$.

Writing Eq.~\eqref{eq:nm_friedmann} as
$H^{2}(z)=H_0^{2}\left[\Omega_m(1+z)^{3}+M(z)\right]$ defines the memory
functional
\begin{equation}
  M(z)\equiv \Omega_\Lambda + \frac{1}{H_0^{2}}\int_{0}^{z} K(z-z')\,\HM{z'}\,dz',
  \label{eq:memory_functional}
\end{equation}
which plays the role of an effective dark-energy density. The associated
effective equation of state is
\begin{equation}
  w_{\rm eff}(z) = -1 + \frac{1}{3}\,\frac{d\ln M(z)}{d\ln(1+z)}.
  \label{eq:weff}
\end{equation}

\section{Five continuous memory kernels}
\label{sec:kernels}

We consider five continuous kernels $K(\Delta z)$ substituted into
Eq.~\eqref{eq:nm_friedmann}. In every case the Markovian (\LCDM) limit is
recovered as the amplitude $A\to0$.

\paragraph*{Linear kernel.} The memory weight grows in proportion to the
redshift lag, with a single free amplitude $A$:
\begin{equation}
  H^{2}(z) = \HM{z} + \int_{0}^{z} A\,(z-z')\,\HM{z'}\,dz'.
  \label{eq:linear}
\end{equation}

\paragraph*{Power-law kernel.} A scale-free algebraic decay of memory, where
$\beta>0$ sets the decay rate and larger $\beta$ suppresses contributions from
the distant past:
\begin{equation}
  H^{2}(z) = \HM{z} + \int_{0}^{z} \frac{A}{(1+z-z')^{\beta}}\,\HM{z'}\,dz'.
  \label{eq:powerlaw}
\end{equation}

\paragraph*{Exponential kernel.} Rapidly fading memory with decay rate
$\alpha>0$:
\begin{equation}
  H^{2}(z) = \HM{z} + \int_{0}^{z} A\,e^{-\alpha(z-z')}\,\HM{z'}\,dz'.
  \label{eq:exponential}
\end{equation}

\paragraph*{Gaussian kernel.} A smooth, symmetric decay whose width $\sigma$
sets the redshift scale over which memory remains significant:
\begin{equation}
  H^{2}(z) = \HM{z} + \int_{0}^{z} A\,\exp\!\left[-\frac{(z-z')^{2}}{2\sigma^{2}}\right]\HM{z'}\,dz'.
  \label{eq:gaussian}
\end{equation}

\paragraph*{Oscillatory kernel.} Memory that alternates between positive and
negative contributions; the damping rate $\alpha$ controls the overall decay
and the frequency $\omega$ the oscillatory structure:
\begin{equation}
  H^{2}(z) = \HM{z} + A\!\int_{0}^{z} e^{-\alpha(z-z')}\cos[\omega(z-z')]\,\HM{z'}\,dz'.
  \label{eq:oscillatory}
\end{equation}

\paragraph*{Closed-form evaluation.} Because $\HM{z'}/H_0^{2}$ is a cubic
polynomial in $u\equiv z-z'$,
$\Omega_m(1+z-u)^{3}+\Omega_\Lambda=\sum_{k=0}^{3}c_k(z)\,u^{k}$ with
$c_0=\Omega_m(1+z)^{3}+\Omega_\Lambda$, $c_1=-3\Omega_m(1+z)^{2}$,
$c_2=3\Omega_m(1+z)$ and $c_3=-\Omega_m$, the memory integral reduces to
\begin{equation}
  \frac{1}{H_0^{2}}\int_{0}^{z}K(z-z')\,\HM{z'}\,dz'=\sum_{k=0}^{3}c_k(z)\,\mathcal{M}_k(z),
  \label{eq:moments}
\end{equation}
where $\mathcal{M}_k(z)=\int_0^{z}K(u)\,u^{k}\,du$ are the kernel moments. For
all five kernels these moments are elementary or incomplete-gamma functions
(the Oscillatory case follows from the Exponential one with
$\alpha\to\alpha-i\omega$), so Eq.~\eqref{eq:nm_friedmann} is evaluated exactly,
including for very sharply peaked kernels. We verified the closed forms against
direct numerical quadrature to a relative precision of $10^{-14}$.

\section{Physical origin of the memory kernels}
\label{sec:origin}

The kernels above were introduced on mathematical grounds and may therefore
appear purely phenomenological. We now show that each has a physically
grounded microscopic origin, so that the parameters
$(A,\alpha,\beta,\sigma,\omega)$ carry physical meaning rather than serving as
arbitrary shape parameters.

\subsection{Open-system derivation: the kernel as an environmental spectral density}
\label{sec:opensystem}

The coarse-grained FLRW description is not a closed dynamical system:
short-wavelength metric and matter fluctuations, sub-horizon quantum modes and
possible hidden-sector fields have all been eliminated to arrive at
Eq.~\eqref{eq:friedmann}. Whenever a macroscopic variable couples linearly to a
continuum of microscopic degrees of freedom, eliminating the latter through the
Zwanzig--Mori projection-operator formalism~\cite{Zwanzig1961,Mori1965}
necessarily introduces a memory term into the macroscopic equation of motion,
in close analogy with the generalised Langevin equation of quantum Brownian
motion~\cite{Caldeira1983a,Caldeira1983b,Weiss2012}.

Coupling the Ricci scalar $R(t)$ bilinearly to a continuum of bath oscillators
$\{q_i,\omega_i\}$ with couplings $c_i$, and eliminating the bath with retarded
boundary conditions, yields the memory kernel
\begin{equation}
  K(\tau) = \frac{2}{\pi}\int_{0}^{\infty} d\omega\,\frac{J(\omega)}{\omega}\cos(\omega\tau),
  \qquad \tau\leftrightarrow\Delta z,
  \label{eq:kernel_spectral}
\end{equation}
where
$J(\omega)=\frac{\pi}{2}\sum_i \frac{c_i^{2}}{m_i\omega_i}\,\delta(\omega-\omega_i)$
is the bath spectral density~\cite{Kubo1966}. Equation~\eqref{eq:kernel_spectral}
is exact in proper time $\tau=t-t'$. Since $dz=-(1+z)H(z)\,dt$, the
identification $\tau\leftrightarrow\Delta z$ used in
Eqs.~\eqref{eq:linear}--\eqref{eq:oscillatory} is accurate where $H(z)$ varies
slowly across the fitted range, and becomes exact in e-fold time $N=\ln a$.
Each of the five kernels corresponds to a canonical choice of $J(\omega)$, as
summarised in Table~\ref{tab:spectral}. Explicitly,
\begin{align}
  J(\omega) &= \eta\,\alpha\,\frac{\omega}{\omega^{2}+\alpha^{2}}
    &&\Rightarrow\quad K(\tau)=A\,e^{-\alpha\tau}, \label{eq:J_exp}\\
  J(\omega) &\propto \omega^{\beta}e^{-\epsilon\omega}
    &&\Rightarrow\quad K(\tau)\xrightarrow{\;\epsilon\to1\;}\frac{A}{(1+\tau)^{\beta}}, \label{eq:J_pow}\\
  J(\omega) &\propto \omega\,e^{-\omega^{2}\sigma^{2}/2}
    &&\Rightarrow\quad K(\tau)=A\,e^{-\tau^{2}/2\sigma^{2}}, \label{eq:J_gauss}
\end{align}
\begin{multline}
  J(\omega) \propto \omega\left[\frac{\alpha}{(\omega-\omega_0)^{2}+\alpha^{2}}
      +\frac{\alpha}{(\omega+\omega_0)^{2}+\alpha^{2}}\right]\\
  \Rightarrow\quad K(\tau)=A\,e^{-\alpha\tau}\cos(\omega_0\tau). \label{eq:J_osc}
\end{multline}
In Eqs.~\eqref{eq:J_exp} and \eqref{eq:J_osc} the parameter $\alpha$ is the
width of a Lorentzian and is therefore positive: a physical bath always
produces a decaying memory envelope. The $\omega_0\to0$ limit of
Eq.~\eqref{eq:J_osc} reduces identically to Eq.~\eqref{eq:J_exp}. The Linear
kernel emerges as the $\tau\to0$ Taylor limit of any spectral density with
$J(0)=0$; it is therefore the model-independent leading-order truncation of any
microscopic realisation, rather than a sixth, independent model.

\paragraph*{Short-memory limit.} When the kernel is sharply peaked compared with
the redshift range probed ($\alpha,\beta\gg1$ or $\sigma\ll1$), it acts as
$K(\Delta z)\simeq\Aeff\,\delta(\Delta z)$ with integrated amplitude
\begin{equation}
  \Aeff\equiv\int_0^{\infty}K(u)\,du=
  \begin{cases}
    A/\alpha & \text{Exponential},\\
    A/(\beta-1) & \text{Power law},\\
    A\sigma\sqrt{\pi/2} & \text{Gaussian},\\
    A\alpha/(\alpha^{2}+\omega^{2}) & \text{Oscillatory},
  \end{cases}
  \label{eq:Aeff}
\end{equation}
and Eq.~\eqref{eq:nm_friedmann} becomes
\begin{equation}
  H^{2}(z)\simeq(1+\Aeff)\,\HM{z}.
  \label{eq:shortmemory}
\end{equation}
In this limit the memory is indistinguishable from a rescaling of the Hubble
constant, $H_0\to H_0\sqrt{1+\Aeff}$, and the shape of the kernel is lost. This
explains the observational degeneracy found in Sec.~\ref{sec:fits}.

\begin{table*}[!t]
  \caption{Spectral densities $J(\omega)$ that generate each memory kernel
  through Eq.~\eqref{eq:kernel_spectral}, together with their physical
  interpretation.}
  \label{tab:spectral}
  \begin{ruledtabular}
  \begin{tabular}{llll}
    Kernel & Spectral density $J(\omega)$ & Kernel $K(\tau)$ & Physical interpretation \\
    \hline
    Exponential & $\eta\alpha\,\omega/(\omega^{2}+\alpha^{2})$ (Drude-regularised Ohmic)
                & $A\,e^{-\alpha\tau}$
                & Hidden continuum with correlation time $\alpha^{-1}$ \\
    Power law   & $\omega^{\beta}e^{-\epsilon\omega}$
                & $A\,(1+\tau)^{-\beta}$
                & Scale-free hidden sector \\
    Gaussian    & $\omega\,e^{-\omega^{2}\sigma^{2}/2}$
                & $A\,e^{-\tau^{2}/2\sigma^{2}}$
                & Band-limited hidden sector \\
    Oscillatory & Lorentzian doublet at $\pm\omega_0$, width $\alpha>0$
                & $A\,e^{-\alpha\tau}\cos(\omega_0\tau)$
                & Single damped hidden resonance \\
    Linear      & $\tau\to0$ limit of any bath with $J(0)=0$
                & $A\,\tau$
                & Universal leading-order truncation \\
  \end{tabular}
  \end{ruledtabular}
\end{table*}

\subsection{Causal thermodynamics: Israel--Stewart derivation of the exponential kernel}
\label{sec:israel}

The exponential kernel can be derived independently by treating the dark
sector as an imperfect fluid with causal bulk-viscous stress. The non-causal
Eckart closure $\Pi=-3\zeta H$ is known to permit superluminal propagation and
generic instabilities~\cite{Hiscock1983,Hiscock1985}. The causal Israel--Stewart
theory~\cite{Israel1976,IsraelStewart1979} instead imposes the relaxation
equation
\begin{equation}
  \tau\,\dot{\Pi} + \Pi = -3\zeta H,
  \label{eq:IS}
\end{equation}
whose late-time (attractor) solution is
\begin{equation}
  \Pi(t) = -\frac{3\zeta}{\tau}\int_{-\infty}^{t} dt'\,e^{-(t-t')/\tau}\,H(t').
  \label{eq:IS_solution}
\end{equation}
Inserting Eq.~\eqref{eq:IS_solution} into the energy-conservation equation and
linearising about the background solution reproduces the Exponential kernel,
Eq.~\eqref{eq:exponential}, in the fast-relaxation limit $\alpha\equiv1/\tau\gg H_0$,
with $A\propto\zeta/\tau$. The parameters $A$ and $\alpha=1/\tau$ are then
identified as transport coefficients of the cosmic fluid, and the integrated
amplitude of Eq.~\eqref{eq:Aeff} becomes $\Aeff=A/\alpha\propto\zeta$, the bulk
viscosity itself. The best fits of Sec.~\ref{sec:fits} lie in the
fast-relaxation regime ($\alpha\simeq26$ for CC, $16$ for DR1 and at the prior
boundary $\alpha=100$ for DR2). In this regime the data constrain $\zeta$
through $\Aeff$, but not the relaxation time $\tau$ separately.

\subsection{Non-local gravity: the Deser--Woodard retarded Green's function}
\label{sec:deserwoodard}

An independent route arises from non-local modifications of the gravitational
action,
\begin{equation}
  S = \frac{1}{2\kappa}\int d^{4}x\,\sqrt{-g}\,\left[R + R\,f\!\left(\Box^{-1}R\right)\right],
  \label{eq:DW_action}
\end{equation}
where $\Box^{-1}$ is defined through the retarded Green's function to preserve
causality~\cite{Deser2007}. On an FLRW background~\cite{Deffayet2009},
\begin{equation}
  \Box^{-1}R(t) = -\int_{-\infty}^{t} dt'\,G(t,t')\,R(t'),
  \label{eq:DW_green}
\end{equation}
with
\begin{equation}
  G(t,t') = a^{3}(t')\int_{t'}^{t}\frac{dt''}{a^{3}(t'')}.
  \label{eq:DW_G}
\end{equation}
During matter or radiation domination, expanding Eq.~\eqref{eq:DW_G} for a
small lag $\Delta t\ll t$ gives $G(\Delta t)\simeq\Delta t$, recovering the
Linear kernel as the short-lag limit of the theory's own retarded propagator.
In de~Sitter domination, $a=a_0e^{H_\Lambda t}$, Eq.~\eqref{eq:DW_G} is
exactly solvable:
\begin{equation}
  G(\Delta t) = \frac{1}{3H_\Lambda}\left(1-e^{-3H_\Lambda\Delta t}\right).
  \label{eq:DW_deSitter}
\end{equation}
This is a saturating rather than a decaying exponential kernel, in line with
the known late-time behaviour of non-local dark
energy~\cite{Nojiri2008,Maggiore2014,Dirian2015,Dirian2016}. We note that the
Power-law and Gaussian kernels cannot be derived independently from non-local
gravity; for these we rely on the open-system derivation of
Sec.~\ref{sec:opensystem}.

\section{Observational data}
\label{sec:data}

We use Hubble-parameter measurements from Cosmic Chronometers and BAO
measurements from the Dark Energy Spectroscopic Instrument (DESI).

\paragraph*{Cosmic Chronometers.} We use 30 model-independent $H(z)$
measurements spanning $0.07\le z\le1.965$, compiled from
Refs.~\cite{Simon2005,Stern2010,Moresco2012a,Zhang2014,Moresco2015,Moresco2016,Ratsimbazafy2017}.
The Hubble parameter is estimated through the differential-age
method~\cite{Jimenez2002}, $H(z)=-(1+z)^{-1}\,dz/dt$, which provides
model-independent constraints on the expansion
history~\cite{Moresco2012b}.

\paragraph*{DESI BAO.} DESI measures the Hubble distance $D_H(z)=c/H(z)$ in
units of the sound horizon at the drag epoch. We convert the published $D_H/r_d$
values and their $1\sigma$ errors into $H(z)$ through
\begin{equation}
  H(z)=\frac{c}{(D_H/r_d)\,r_d},\qquad
  \sigma_H=H\,\frac{\sigma_{D_H/r_d}}{D_H/r_d},
  \label{eq:dh_to_h}
\end{equation}
with $r_d=147.09$~Mpc. For DR1~\cite{DESI2024,DESI2024III,DESI2024IV} this
yields five points from the LRG1, LRG2, LRG3+ELG1, ELG2 and Ly$\alpha$ samples;
the BGS and QSO samples provide only the isotropic $D_V/r_d$ and are not used.
For DR2~\cite{DESI2025,DESI2025Lya} the QSO sample also provides $D_H/r_d$,
giving six points. The resulting values are listed in Table~\ref{tab:desi}.
DR1 and DR2 overlap in sky coverage and tracers, so they are never combined;
we analyse CC, DR1, DR2, CC+DR1 and CC+DR2.

\begin{table}[t]
  \caption{DESI $H(z)$ values derived from $D_H/r_d$ through
  Eq.~\eqref{eq:dh_to_h} with $r_d=147.09$~Mpc. DR1 data are from
  Refs.~\cite{DESI2024,DESI2024III,DESI2024IV}; DR2 data are from
  Refs.~\cite{DESI2025,DESI2025Lya}.}
  \label{tab:desi}
  \begin{ruledtabular}
  \begin{tabular}{lcc}
    Release & $z_{\rm eff}$ & $H(z)$ [\kms] \\
    \hline
    DR1 & 0.510 & $97.15\pm2.82$ \\
        & 0.706 & $101.50\pm3.03$ \\
        & 0.930 & $113.99\pm2.23$ \\
        & 1.317 & $147.48\pm4.48$ \\
        & 2.330 & $239.22\pm4.77$ \\
    \hline
    DR2 & 0.510 & $93.22\pm1.81$ \\
        & 0.706 & $104.76\pm1.78$ \\
        & 0.934 & $115.54\pm1.26$ \\
        & 1.321 & $143.78\pm2.24$ \\
        & 1.484 & $159.02\pm6.40$ \\
        & 2.330 & $236.12\pm2.76$ \\
  \end{tabular}
  \end{ruledtabular}
\end{table}

\section{Statistical method}
\label{sec:method}

The agreement between a model and a dataset is quantified by
\begin{equation}
  \chi^{2}(\boldsymbol{\theta}) = \sum_{i=1}^{N}
  \left[\frac{H_{\rm obs}(z_i)-H_{\rm model}(z_i;\boldsymbol{\theta})}{\sigma_i}\right]^{2},
  \label{eq:chi2}
\end{equation}
where $H_{\rm obs}(z_i)$ and $\sigma_i$ are the measured Hubble parameter and its
$1\sigma$ uncertainty at redshift $z_i$, $N$ is the number of data points, and
$\boldsymbol{\theta}$ denotes the free kernel parameters ($A$, $\alpha$,
$\beta$, $\sigma$, $\omega$ as appropriate). The background parameters are held
at their Planck 2018 values, so the Markovian (\LCDM) model has no free
parameters.

For each kernel and dataset we first minimise Eq.~\eqref{eq:chi2} with the
Nelder--Mead algorithm, started from one physically motivated point and eight
random points in the prior volume to avoid local minima. We then sample the
posterior $\mathcal{P}\propto e^{-\chi^{2}/2}$ with the affine-invariant
ensemble sampler \texttt{emcee}~\cite{emcee}, using 32 walkers and 6000 steps,
discarding the first 1500 steps as burn-in and thinning by 5, which leaves
28\,800 samples per run. The walkers are initialised in a small ball around the
Nelder--Mead minimum, and the reported $\chi^{2}_{\rm min}$ is the lower of the
optimiser and chain minima. Flat priors are adopted:
$A\in(-2,10)$ for the Linear kernel, $A\in(-2,50)$ otherwise,
$\beta\in(0.1,100)$, $\alpha\in(0.01,100)$ for the Exponential kernel,
$\alpha\in(-2,100)$ for the Oscillatory kernel, $\sigma\in(0.01,5)$ and
$\omega\in(0,20)$. The Oscillatory prior deliberately admits $\alpha<0$ in
order to test whether the data demand a growing envelope. Posterior summaries
are medians with 16th and 84th percentiles; contours are produced with
\texttt{GetDist}~\cite{getdist}. Integrated autocorrelation times are
$\tau_{\rm int}\simeq26$--$73$ steps for all runs except the Gaussian kernel
($\tau_{\rm int}\lesssim166$), with mean acceptance fractions of 0.48--0.81.

Models are compared through
\begin{equation}
  {\rm AIC}=\chi^{2}_{\rm min}+2k,\qquad {\rm BIC}=\chi^{2}_{\rm min}+k\ln N,
  \label{eq:ic}
\end{equation}
where $k$ is the number of free parameters ($k=0$ for \LCDM). We quote
$\dchi\equiv\chi^{2}_{\Lambda{\rm CDM}}-\chi^{2}_{\rm NM}$, so positive values
favour the non-Markovian model, and $\Delta{\rm AIC}={\rm AIC}_{\rm NM}-{\rm
AIC}_{\Lambda{\rm CDM}}$ and $\Delta{\rm BIC}$ likewise, so negative values
favour the non-Markovian model. Differences of 2, 6 and 10 are conventionally
read as positive, strong and very strong evidence.

\section{Results}
\label{sec:fits}

\subsection{Markovian baseline}

With the Planck 2018 background, \LCDM{} gives $\chi^{2}=14.894$ for CC
($N=30$), $8.401$ for DR1 ($N=5$), $8.626$ for DR2 ($N=6$), $23.295$ for CC+DR1
and $23.521$ for CC+DR2. \LCDM{} therefore already provides an acceptable
description of all five datasets. The best-fit parameters of the five kernels
are collected in Table~\ref{tab:bestfit}, the model comparison in
Table~\ref{tab:ic}, and the MCMC constraints in Table~\ref{tab:posterior}.

\begin{table*}[!t]
  \caption{Best-fit kernel parameters and minimum $\chi^{2}$ for each dataset,
  from multi-start Nelder--Mead minimisation refined by MCMC. The first row of
  each block gives the Markovian (\LCDM) reference. The second-parameter column
  lists $\beta$, $\alpha$ or $\sigma$ as appropriate; values marked $^{\dagger}$
  lie on the prior boundary. The last column gives the integrated amplitude
  $\Aeff$ of Eq.~\eqref{eq:Aeff} for the three degenerate kernels.}
  \label{tab:bestfit}
  \begin{ruledtabular}
  \begin{tabular}{lllllr}
    Kernel & $A$ & Second parameter & $\omega$ & $\Aeff$ & $\chi^{2}_{\rm min}$ \\
    \hline
    \multicolumn{6}{l}{\textit{Cosmic Chronometers} ($N=30$, $0.07\le z\le1.965$)}\\
    Markovian (\LCDM) & --- & --- & --- & --- & 14.894 \\
    Linear      & 0.1199 & ---                         & ---   & ---    & 14.675 \\
    Power law   & 0.8538 & $\beta=29.40$               & ---   & 0.0301 & 14.519 \\
    Exponential & 0.7797 & $\alpha=25.85$              & ---   & 0.0302 & 14.519 \\
    Gaussian    & 0.3507 & $\sigma=0.0698$             & ---   & 0.0307 & 14.515 \\
    Oscillatory & 1.302  & $\alpha=-2.00^{\dagger}$    & 14.76 & ---    &  9.099 \\
    \hline
    \multicolumn{6}{l}{\textit{DESI DR1} ($N=5$, $0.51\le z\le2.33$)}\\
    Markovian (\LCDM) & --- & --- & --- & --- & 8.401 \\
    Linear      & 0.0371 & ---                         & ---   & ---    & 7.948 \\
    Power law   & 0.5004 & $\beta=21.31$               & ---   & 0.0246 & 7.284 \\
    Exponential & 0.3904 & $\alpha=15.64$              & ---   & 0.0250 & 7.284 \\
    Gaussian    & 0.1407 & $\sigma=0.150$              & ---   & 0.0265 & 7.282 \\
    Oscillatory & 1.319  & $\alpha=-0.099$             & 5.011 & ---    & 0.332 \\
    \hline
    \multicolumn{6}{l}{\textit{DESI DR2} ($N=6$, $0.51\le z\le2.33$)}\\
    Markovian (\LCDM) & --- & --- & --- & --- & 8.626 \\
    Linear      & $-0.0003$ & ---                      & ---   & ---    & 8.626 \\
    Power law   & 1.037  & $\beta=100^{\dagger}$       & ---   & 0.0105 & 7.926 \\
    Exponential & 1.048  & $\alpha=100^{\dagger}$      & ---   & 0.0105 & 7.926 \\
    Gaussian    & 0.8344 & $\sigma=0.01^{\dagger}$     & ---   & 0.0105 & 7.925 \\
    Oscillatory & 0.5026 & $\alpha=-0.447$             & 4.170 & ---    & 1.007 \\
    \hline
    \multicolumn{6}{l}{\textit{CC+DR1} ($N=35$, $0.07\le z\le2.33$)}\\
    Markovian (\LCDM) & --- & --- & --- & --- & 23.295 \\
    Linear      & 0.0408 & ---                         & ---   & ---    & 22.722 \\
    Power law   & 0.7647 & $\beta=31.32$               & ---   & 0.0252 & 21.815 \\
    Exponential & 0.6944 & $\alpha=27.46$              & ---   & 0.0253 & 21.815 \\
    Gaussian    & 0.3058 & $\sigma=0.0673$             & ---   & 0.0258 & 21.811 \\
    Oscillatory & 0.6765 & $\alpha=-0.020$             & 5.211 & ---    & 18.774 \\
    \hline
    \multicolumn{6}{l}{\textit{CC+DR2} ($N=36$, $0.07\le z\le2.33$)}\\
    Markovian (\LCDM) & --- & --- & --- & --- & 23.521 \\
    Linear      & 0.0015 & ---                         & ---   & ---    & 23.518 \\
    Power law   & 1.157  & $\beta=100^{\dagger}$       & ---   & 0.0117 & 22.588 \\
    Exponential & 1.168  & $\alpha=100^{\dagger}$      & ---   & 0.0117 & 22.588 \\
    Gaussian    & 0.9305 & $\sigma=0.01^{\dagger}$     & ---   & 0.0117 & 22.588 \\
    Oscillatory & 0.3735 & $\alpha=-0.421$             & 4.126 & ---    & 17.724 \\
  \end{tabular}
  \end{ruledtabular}
\end{table*}

\begin{table*}[!t]
  \caption{Model comparison against \LCDM{}. $\dchi>0$ means a better fit;
  $\Delta{\rm AIC}<0$ and $\Delta{\rm BIC}<0$ mean the kernel is preferred after
  penalising its $k$ free parameters.}
  \label{tab:ic}
  \begin{ruledtabular}
  \begin{tabular}{llrrrrr}
    & Kernel ($k$) & CC & DESI DR1 & DESI DR2 & CC+DR1 & CC+DR2 \\
    \hline
    \multirow{5}{*}{$\dchi$}
      & Linear (1)      & $+0.22$ & $+0.45$ & $+0.00$ & $+0.57$ & $+0.00$ \\
      & Power law (2)   & $+0.37$ & $+1.12$ & $+0.70$ & $+1.48$ & $+0.93$ \\
      & Exponential (2) & $+0.38$ & $+1.12$ & $+0.70$ & $+1.48$ & $+0.93$ \\
      & Gaussian (2)    & $+0.38$ & $+1.12$ & $+0.70$ & $+1.48$ & $+0.93$ \\
      & Oscillatory (3) & $+5.80$ & $+8.07$ & $+7.62$ & $+4.52$ & $+5.80$ \\
    \hline
    \multirow{5}{*}{$\Delta{\rm AIC}$}
      & Linear (1)      & $+1.78$ & $+1.55$ & $+2.00$ & $+1.43$ & $+2.00$ \\
      & Power law (2)   & $+3.63$ & $+2.88$ & $+3.30$ & $+2.52$ & $+3.07$ \\
      & Exponential (2) & $+3.62$ & $+2.88$ & $+3.30$ & $+2.52$ & $+3.07$ \\
      & Gaussian (2)    & $+3.62$ & $+2.88$ & $+3.30$ & $+2.52$ & $+3.07$ \\
      & Oscillatory (3) & $+0.20$ & $-2.07$ & $-1.62$ & $+1.48$ & $+0.20$ \\
    \hline
    \multirow{5}{*}{$\Delta{\rm BIC}$}
      & Linear (1)      & $+3.18$ & $+1.16$ & $+1.79$ & $+2.98$ & $+3.58$ \\
      & Power law (2)   & $+6.43$ & $+2.10$ & $+2.88$ & $+5.63$ & $+6.23$ \\
      & Exponential (2) & $+6.43$ & $+2.10$ & $+2.88$ & $+5.63$ & $+6.23$ \\
      & Gaussian (2)    & $+6.42$ & $+2.10$ & $+2.88$ & $+5.63$ & $+6.23$ \\
      & Oscillatory (3) & $+4.41$ & $-3.24$ & $-2.24$ & $+6.15$ & $+4.95$ \\
  \end{tabular}
  \end{ruledtabular}
\end{table*}

\subsection{MCMC posterior constraints}

The marginalised posterior constraints for every kernel and dataset are listed
in Table~\ref{tab:posterior}, and the one- and two-dimensional posteriors from
the individual datasets are shown in Figs.~\ref{fig:tri_lin}--\ref{fig:tri_osc}.
Parameters whose posterior is bounded by the prior rather than by the data are
marked in the table; for these the quoted interval reflects the prior range and
should not be read as a measurement.

\begin{table*}[!t]
  \caption{Marginalised MCMC constraints (median with 16th and 84th
  percentiles) for each kernel and dataset. Entries marked $^{\ast}$ are
  prior-dominated: their posteriors rise towards, or are flat up to, the prior
  boundary (Figs.~\ref{fig:tri_pow}--\ref{fig:tri_osc}). Priors:
  $\beta,\alpha\in(0.01,100)$, $\alpha_{\rm osc}\in(-2,100)$,
  $\sigma\in(0.01,5)$, $\omega\in(0,20)$.}
  \label{tab:posterior}
  \begin{ruledtabular}
  \begin{tabular}{llccccc}
    Kernel & Parameter & CC & DESI DR1 & DESI DR2 & CC+DR1 & CC+DR2 \\
    \hline
    Linear      & $A$
      & $0.127^{+0.263}_{-0.252}$ & $0.039^{+0.056}_{-0.055}$ & $-0.000^{+0.031}_{-0.031}$
      & $0.042^{+0.054}_{-0.053}$ & $0.002^{+0.031}_{-0.030}$ \\
    \hline
    Power law   & $A$
      & $2.01^{+3.47}_{-2.20}$ & $1.40^{+1.85}_{-1.28}$ & $0.63^{+0.97}_{-0.70}$
      & $1.43^{+1.69}_{-1.20}$ & $0.67^{+0.98}_{-0.68}$ \\
                & $\beta^{\ast}$
      & $67.8^{+22.5}_{-30.9}$ & $69.0^{+22.0}_{-30.5}$ & $71.3^{+20.4}_{-31.3}$
      & $70.6^{+21.2}_{-31.1}$ & $70.6^{+21.0}_{-31.2}$ \\
    \hline
    Exponential & $A$
      & $2.08^{+3.51}_{-2.25}$ & $1.44^{+1.84}_{-1.30}$ & $0.60^{+0.99}_{-0.70}$
      & $1.41^{+1.75}_{-1.20}$ & $0.69^{+1.01}_{-0.70}$ \\
                & $\alpha^{\ast}$
      & $68.7^{+22.5}_{-31.8}$ & $69.9^{+21.6}_{-31.1}$ & $70.0^{+21.7}_{-30.6}$
      & $70.0^{+21.3}_{-31.3}$ & $70.7^{+21.2}_{-31.1}$ \\
    \hline
    Gaussian    & $A$
      & $0.086^{+0.184}_{-0.136}$ & $0.045^{+0.204}_{-0.048}$ & $0.016^{+0.080}_{-0.027}$
      & $0.049^{+0.192}_{-0.047}$ & $0.018^{+0.111}_{-0.028}$ \\
                & $\sigma^{\ast}$
      & $1.87^{+2.12}_{-1.71}$ & $1.32^{+2.46}_{-1.26}$ & $1.22^{+2.53}_{-1.15}$
      & $1.22^{+2.58}_{-1.15}$ & $1.17^{+2.52}_{-1.11}$ \\
    \hline
    Oscillatory & $A$
      & $2.05^{+3.47}_{-2.22}$ & $1.49^{+1.87}_{-1.35}$ & $0.65^{+1.00}_{-0.73}$
      & $1.52^{+1.68}_{-1.24}$ & $0.70^{+0.98}_{-0.71}$ \\
                & $\alpha^{\ast}$
      & $66.1^{+24.3}_{-36.6}$ & $66.9^{+24.0}_{-38.0}$ & $65.0^{+24.9}_{-39.7}$
      & $67.2^{+23.5}_{-38.6}$ & $66.7^{+23.9}_{-39.4}$ \\
                & $\omega^{\ast}$
      & $10.4^{+6.6}_{-7.0}$ & $11.0^{+6.3}_{-7.2}$ & $10.8^{+6.6}_{-7.2}$
      & $10.9^{+6.3}_{-7.4}$ & $10.8^{+6.5}_{-7.2}$ \\
  \end{tabular}
  \end{ruledtabular}
\end{table*}

\subsection{Linear kernel}

\begin{figure}[t]
  \centering
  \figorbox{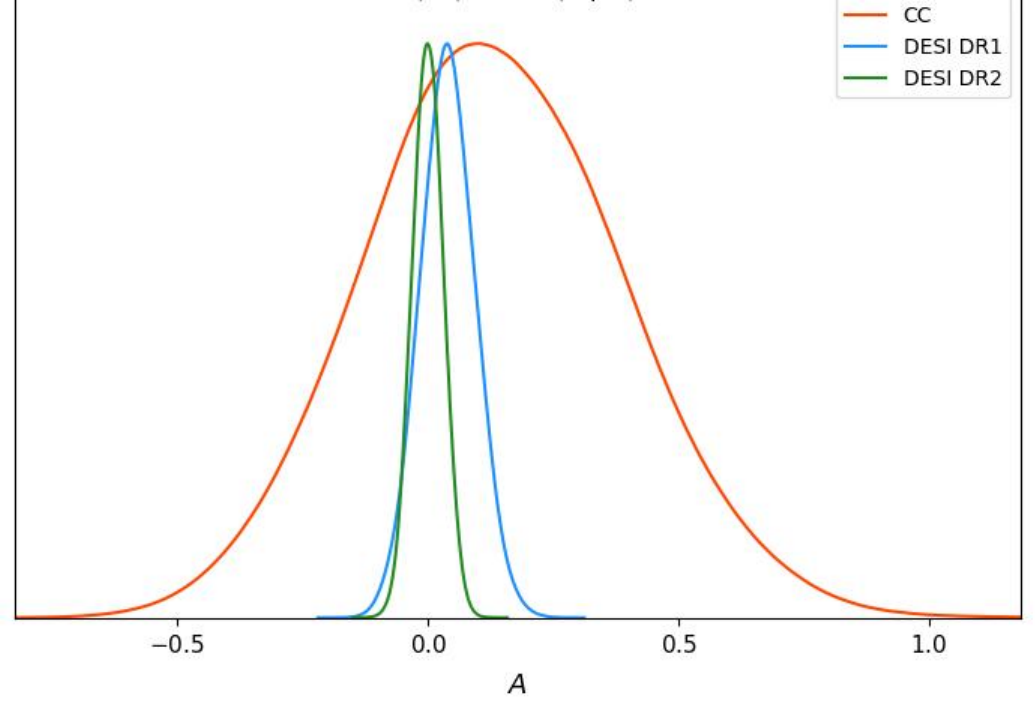}
  \caption{Marginalised posterior of the Linear-kernel amplitude $A$ for CC,
  DESI DR1 and DESI DR2. All three are consistent with the \LCDM{} value $A=0$;
  DR2 gives the tightest bound, $A=-0.000\pm0.031$.}
  \label{fig:tri_lin}
\end{figure}

The Linear kernel improves the fit by at most $\dchi=0.57$, and its amplitude
is consistent with zero in every dataset (Table~\ref{tab:posterior} and
Fig.~\ref{fig:tri_lin}). The posteriors are close to Gaussian and nested: CC
gives a broad constraint, $A=0.127^{+0.263}_{-0.252}$, DR1 tightens it by a
factor of five, and DR2 gives $A=-0.000\pm0.031$, centred on the \LCDM{} value.
Combining CC with DR2 leaves the result essentially unchanged,
$A=0.002\pm0.030$. The DR2 data thus bound the leading-order memory term, which
by Sec.~\ref{sec:opensystem} is common to every microscopic realisation, to
$|A|\lesssim0.06$ at $2\sigma$. The Linear kernel is mildly disfavoured by the
information criteria in all cases ($\Delta{\rm AIC}=+1.4$ to $+2.0$).

\subsection{Power-law, Exponential and Gaussian kernels}

These three kernels give the same $\chi^{2}_{\rm min}$ in every dataset, to
within $0.004$ (Table~\ref{tab:bestfit}). The best fits select sharply peaked
kernels, with $\beta\simeq21$--$31$, $\alpha\simeq16$--$27$ and
$\sigma\simeq0.07$--$0.15$ for CC and DR1, while for DR2 the best fits run to
the prior boundaries. This is the short-memory limit of
Eq.~\eqref{eq:shortmemory}: the three best fits share the same integrated
amplitude, $\Aeff\simeq0.030$ (CC), $0.025$ (DR1), $0.0105$ (DR2), $0.025$
(CC+DR1) and $0.012$ (CC+DR2), and all three simply rescale $H(z)$ by
$\sqrt{1+\Aeff}$. For CC this is equivalent to raising the Planck value of
$H_0$ from 67.4 to $\simeq68.4~\kms$; for DR2, to $\simeq67.8~\kms$.

\begin{figure}[t]
  \centering
  \figorbox{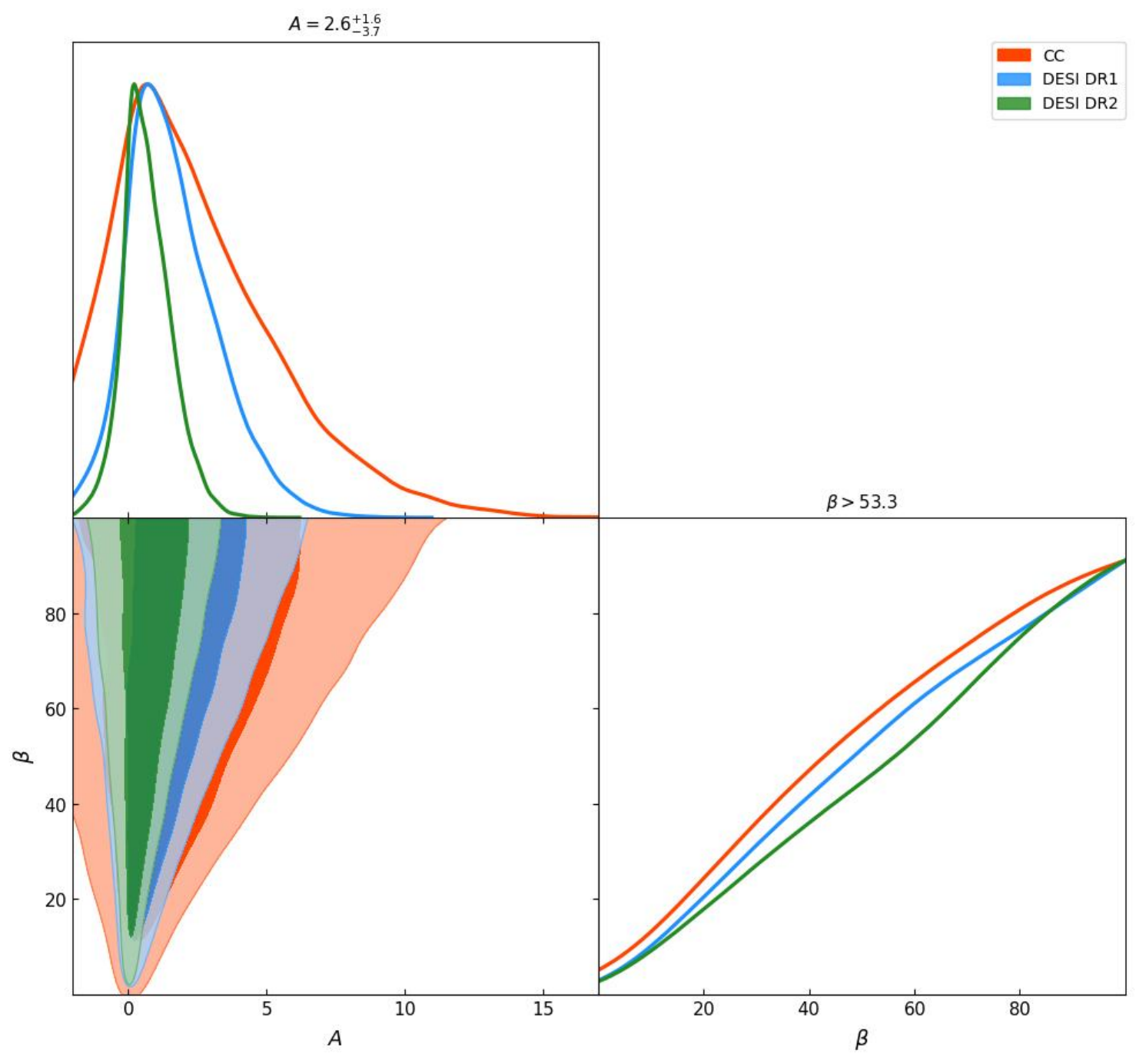}
  \caption{Posteriors (68\% and 95\% contours) for the Power-law kernel from
  CC, DESI DR1 and DESI DR2. The wedge-shaped contours follow the degeneracy
  $A/(\beta-1)\simeq\Aeff$; the marginal $\beta$ posterior rises to the prior
  boundary, giving only a lower limit.}
  \label{fig:tri_pow}
\end{figure}

\begin{figure}[t]
  \centering
  \figorbox{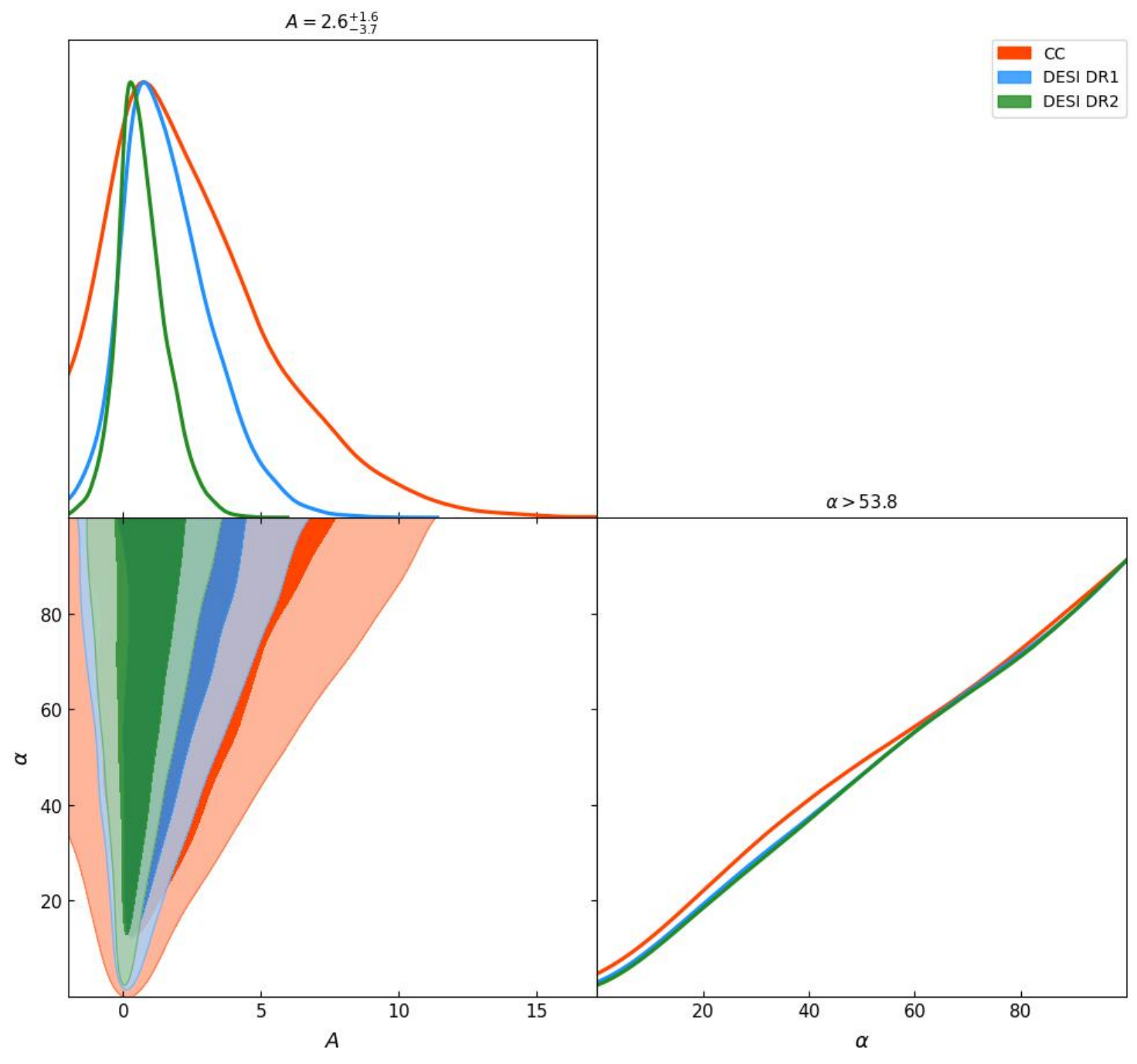}
  \caption{As Fig.~\ref{fig:tri_pow}, for the Exponential kernel. The contours
  follow $A/\alpha\simeq\Aeff$, and the marginal $\alpha$ posterior is bounded
  only by the prior. In the Israel--Stewart interpretation $\Aeff\propto\zeta$ is
  constrained while the relaxation time $\tau=1/\alpha$ is not.}
  \label{fig:tri_exp}
\end{figure}

\begin{figure}[t]
  \centering
  \figorbox{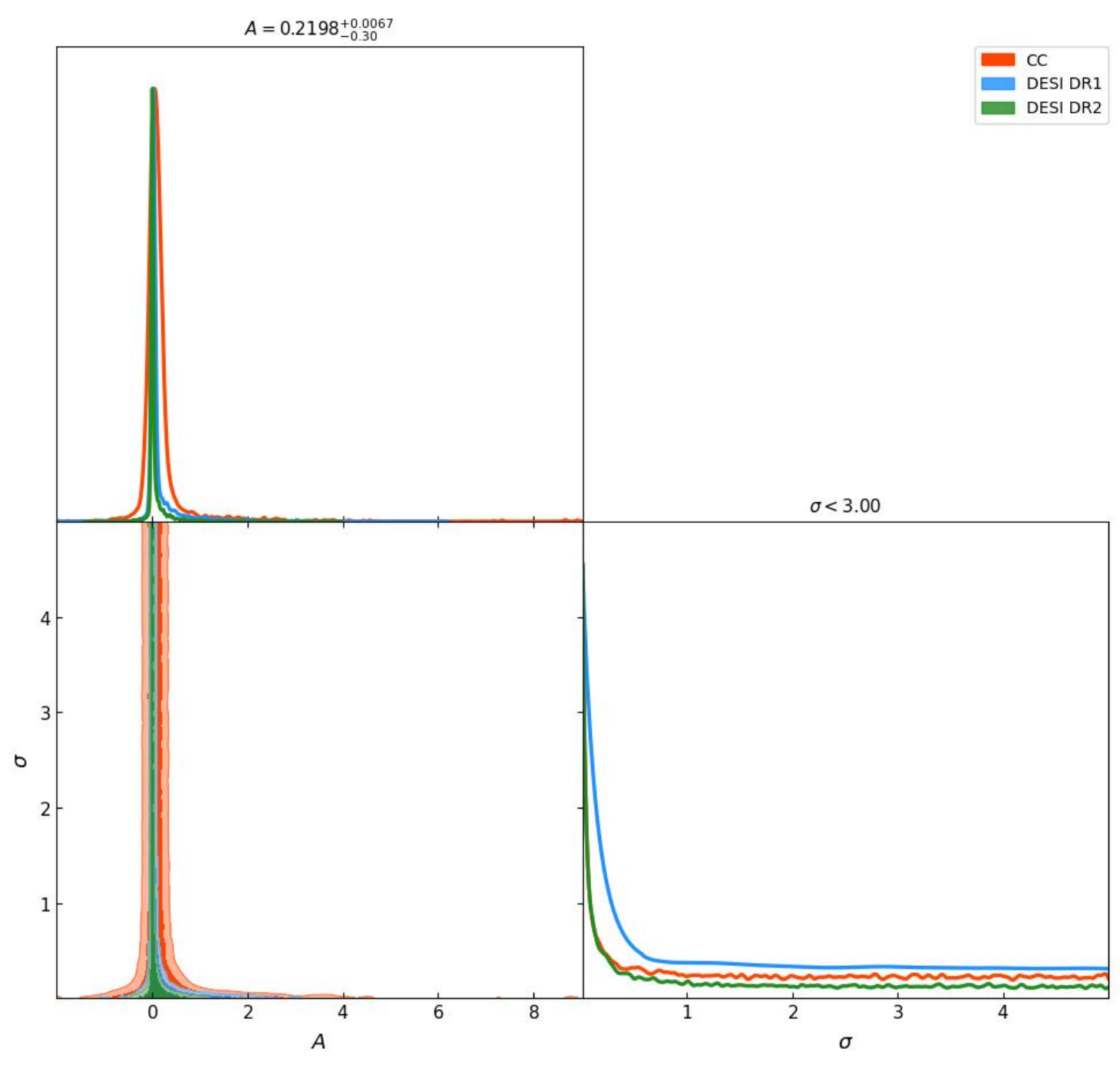}
  \caption{As Fig.~\ref{fig:tri_pow}, for the Gaussian kernel. For
  $\sigma\gtrsim0.5$ the amplitude is confined to a narrow band around $A=0$,
  excluding broad memory; the $\sigma$ posterior is flat once $A\to0$.}
  \label{fig:tri_gauss}
\end{figure}

The MCMC posteriors (Figs.~\ref{fig:tri_pow}--\ref{fig:tri_gauss}) make the
degeneracy explicit. For the Power-law and Exponential kernels the
two-dimensional contours form a wedge that opens towards large $\beta$ or
$\alpha$: when the memory decays slowly ($\alpha,\beta\lesssim10$) the amplitude
is pinned close to zero, while faster decay permits a proportionally larger
amplitude, along the ridge $A/\alpha\simeq A/(\beta-1)\simeq\Aeff$. The
marginal posteriors of $\beta$ and $\alpha$ rise monotonically up to the prior
boundary at 100, so the data provide only a lower limit (at 68\% confidence,
$\beta>53$ and $\alpha>54$ for CC), and the medians in
Table~\ref{tab:posterior}, $\beta\simeq\alpha\simeq70$, are set by the prior.
The amplitude posteriors peak at $A\simeq0.5$--$1$ with a long tail, again
reflecting the ridge. The ordering of the contours is the same in every panel:
DR2 is the tightest, then DR1, then CC.

The Gaussian kernel shows the same physics from the opposite side
(Fig.~\ref{fig:tri_gauss}). The amplitude posterior is sharply peaked at
$A\simeq0$, and for $\sigma\gtrsim0.5$ the contours collapse onto a narrow
vertical band around $A=0$: broad memory, extending over a redshift lag of
order unity, is excluded. The $\sigma$ posterior peaks at the smallest widths
and is flat beyond $\sigma\simeq1$, because once $A\to0$ the width has no
effect on $H(z)$. The Gaussian chains are also the least well mixed
($\tau_{\rm int}\lesssim166$), which shows in the noisier contours.

In all three cases only $\Aeff$ is measured, and the kernel shape and memory
time cannot be recovered from current data. With $\dchi\le1.5$ these kernels
are disfavoured relative to \LCDM{} in every dataset, with
$\Delta{\rm AIC}=+2.5$ to $+3.6$ and $\Delta{\rm BIC}=+2.1$ to $+6.4$.

\subsection{Oscillatory kernel}

The Oscillatory kernel is the only one to improve the fit appreciably,
$\dchi=5.8$ (CC), $8.1$ (DR1), $7.6$ (DR2), $4.5$ (CC+DR1) and $5.8$ (CC+DR2).
The DESI-based fits prefer a common frequency, $\omega\simeq4.1$--$5.2$,
corresponding to a period $\Delta z\simeq1.2$--$1.5$ in redshift lag. However,
every best fit has a negative damping rate, $\alpha=-0.02$ to $-0.45$ for the
DESI-based datasets and $\alpha=-2$ (the prior boundary) for CC, where the fit
uses a high frequency, $\omega=14.8$, to follow scatter in the chronometer
data. A negative $\alpha$ describes a memory kernel that grows into the past.
This is incompatible with the spectral derivation of Eq.~\eqref{eq:J_osc}, in
which $\alpha$ is a Lorentzian width, and with any dissipative interpretation.
The fit to DR1, $\chi^{2}_{\rm min}=0.33$ for three parameters and five data
points, is also a clear sign of over-fitting.

\begin{figure}[t]
  \centering
  \figorbox{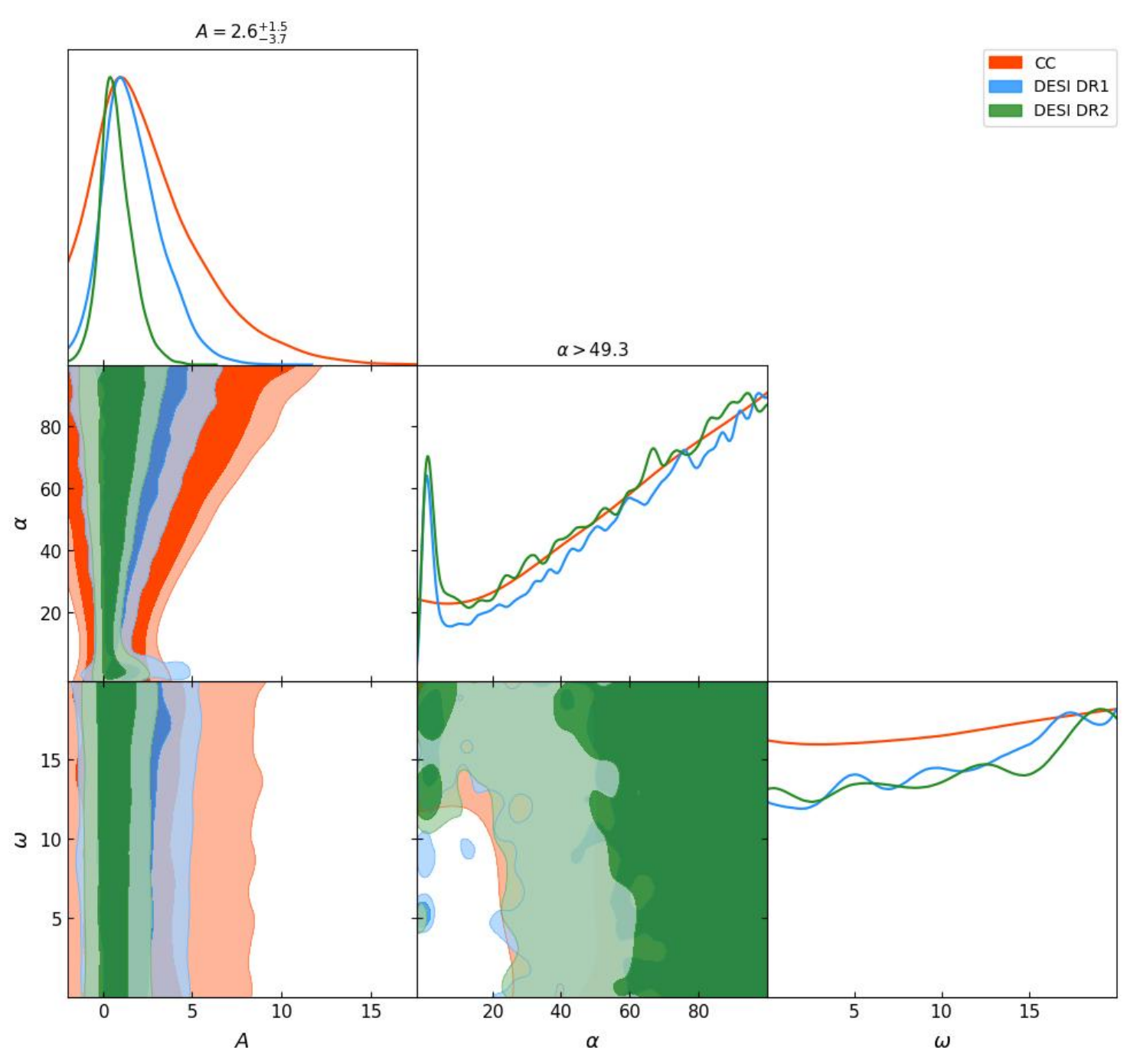}
  \caption{Posteriors (68\% and 95\% contours) for the Oscillatory kernel from
  CC, DESI DR1 and DESI DR2. The frequency $\omega$ is unconstrained; the
  $\alpha$ posterior has a narrow spike near $\alpha\simeq0$ for DR1 and DR2
  (the best-fit oscillating solution) on top of a broad, prior-dominated
  distribution at large $\alpha$.}
  \label{fig:tri_osc}
\end{figure}

The posteriors (Fig.~\ref{fig:tri_osc}) show how little weight this solution
carries. The frequency $\omega$ is unconstrained across its whole prior range,
and the $\alpha$ posterior consists of a narrow spike near $\alpha\simeq0$ for
DR1 and DR2, which is the oscillating solution of Table~\ref{tab:bestfit}, on
top of a broad distribution that rises towards the prior boundary at
$\alpha=100$, where the kernel is effectively a delta function and the model
reverts to the short-memory limit of Eq.~\eqref{eq:shortmemory}. The CC
posterior shows no such spike. Most of the posterior mass therefore lies in the
non-oscillating regime, and the marginal medians ($\alpha\simeq65$--$67$,
$\omega\simeq10$--$11$) are prior-dominated and far from the best fits.

The information criteria reflect this. The Oscillatory kernel is weakly
preferred for DESI alone ($\Delta{\rm AIC}=-2.1$ and $-1.6$, $\Delta{\rm
BIC}=-3.2$ and $-2.2$ for DR1 and DR2), but not for CC or for the joint
datasets ($\Delta{\rm AIC}=+0.2$ to $+1.5$, $\Delta{\rm BIC}=+4.4$ to $+6.2$).

\subsection{Hubble-parameter fits and mean deviation}

\begin{figure}[t]
  \centering
  \figorbox{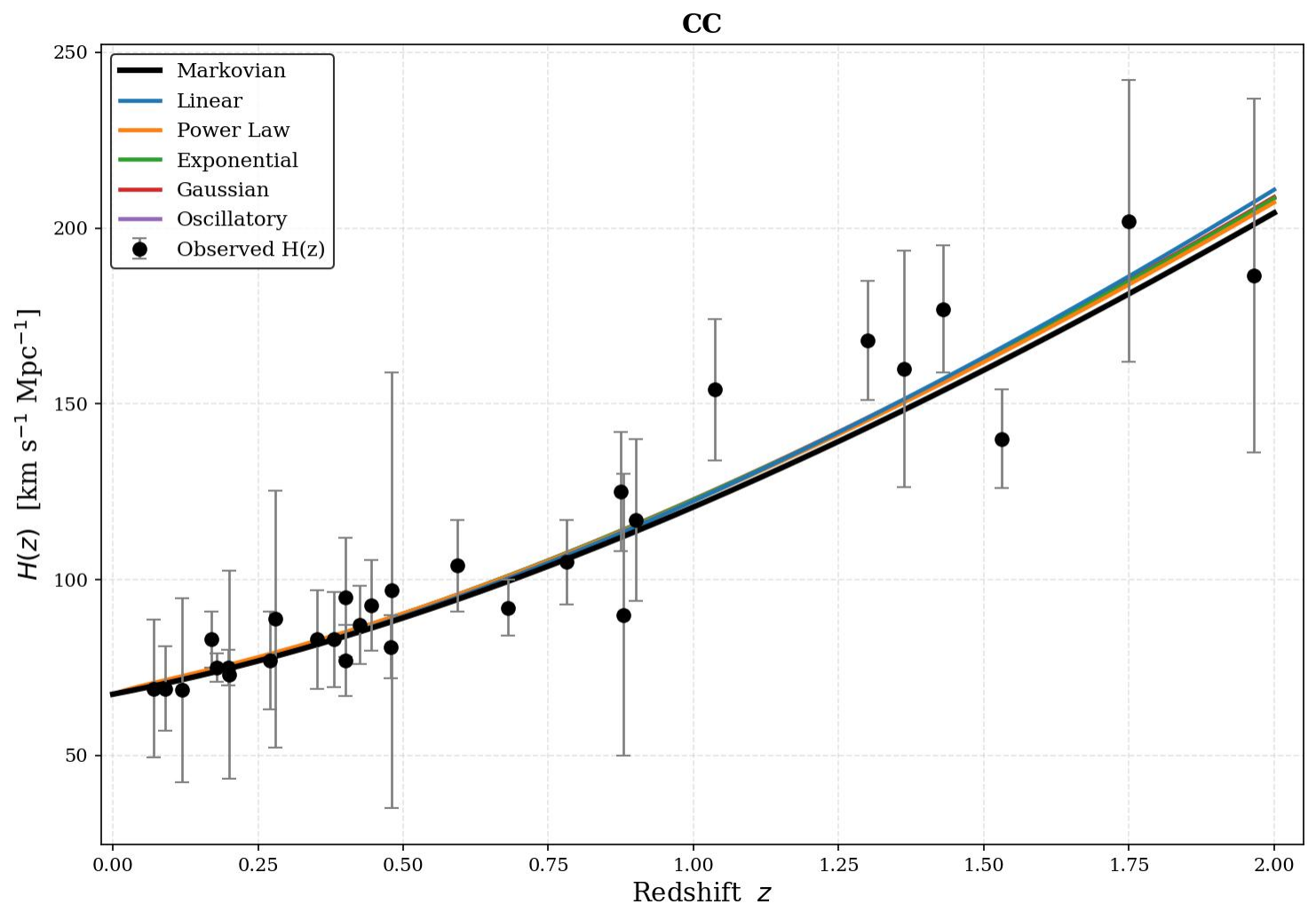}
  \caption{$H(z)$ for the CC dataset (30 points, $0.07\le z\le1.965$). Black
  solid line: Markovian (\LCDM) baseline; coloured lines: the five best-fit
  non-Markovian kernels; black circles: observed data with $1\sigma$ error
  bars.}
  \label{fig:cc}
\end{figure}

\begin{figure}[t]
  \centering
  \figorbox{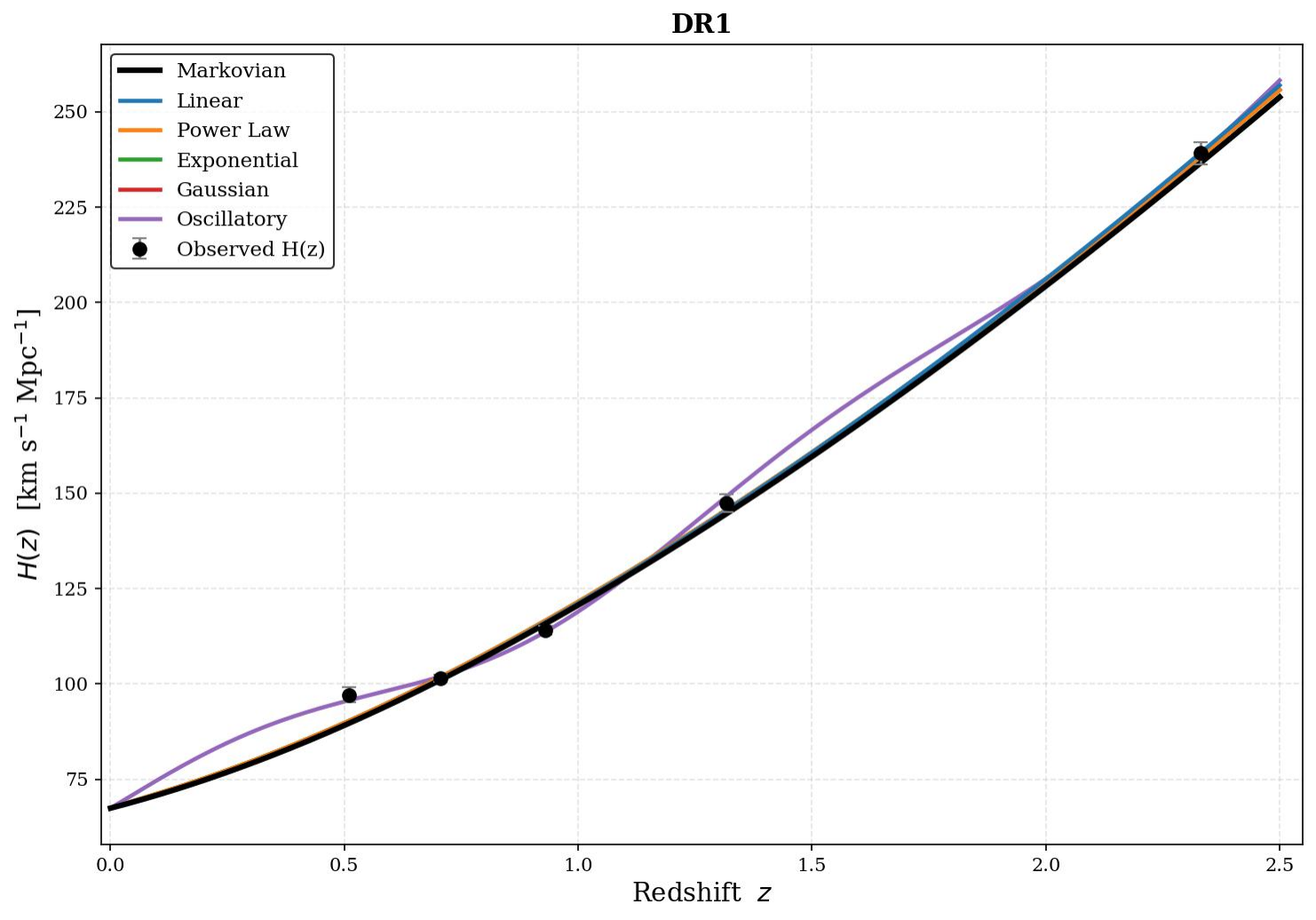}
  \caption{$H(z)$ for DESI DR1 (5 points, $0.51\le z\le2.33$). The Oscillatory
  kernel gives the largest improvement ($\dchi=8.1$) but requires $\alpha<0$.
  Line styles as in Fig.~\ref{fig:cc}.}
  \label{fig:dr1}
\end{figure}

\begin{figure}[t]
  \centering
  \figorbox{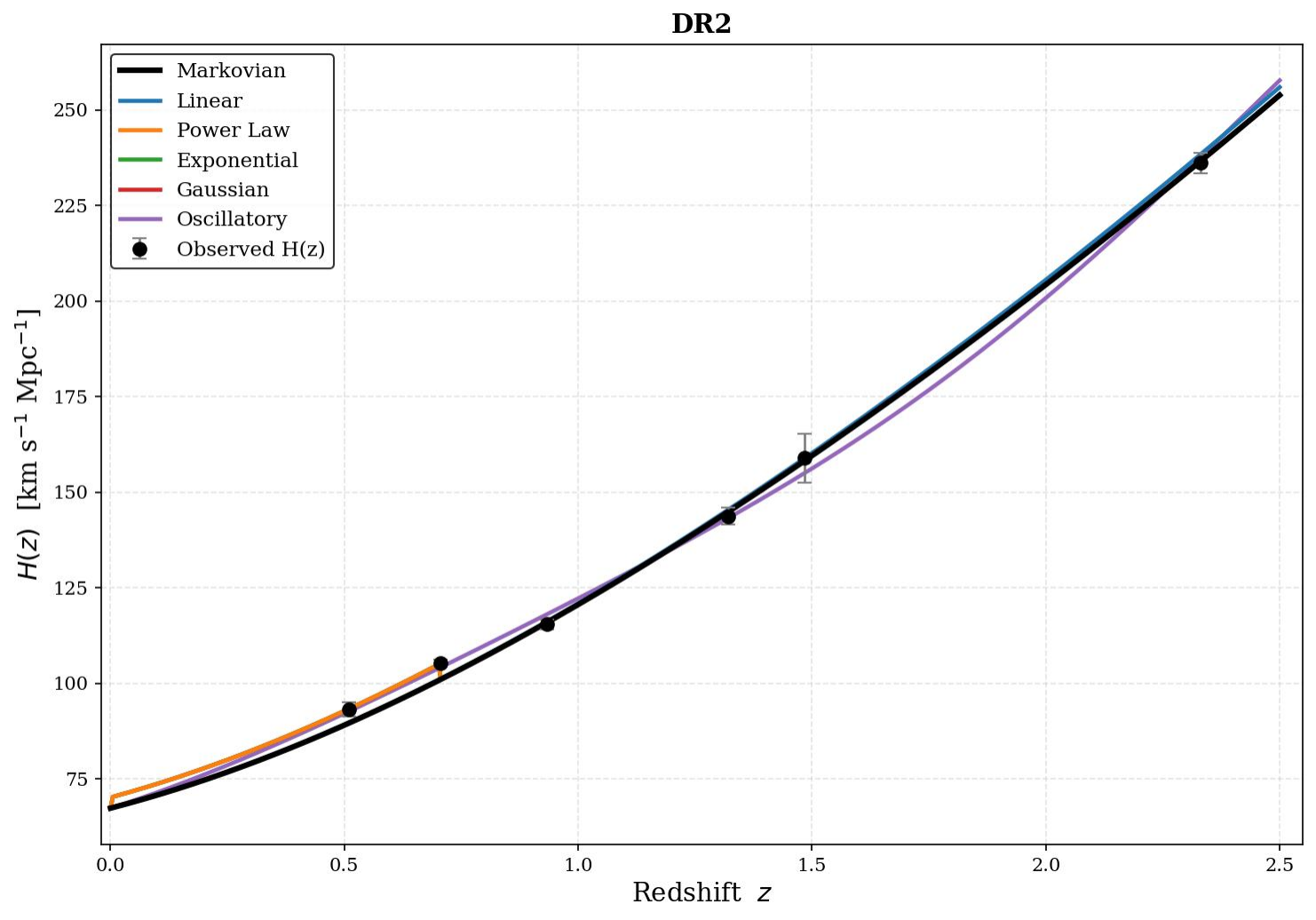}
  \caption{$H(z)$ for DESI DR2 (6 points, $0.51\le z\le2.33$). The Power-law,
  Exponential and Gaussian kernels coincide ($\dchi=0.70$); the Oscillatory
  kernel gives $\dchi=7.6$ with $\alpha<0$. Line styles as in
  Fig.~\ref{fig:cc}.}
  \label{fig:dr2}
\end{figure}

Figures~\ref{fig:cc}--\ref{fig:dr2} compare $H(z)$ for the five best-fit
kernels with the Markovian baseline for the CC, DR1 and DR2 datasets. We
quantify the departure of each kernel from the baseline through the mean
percentage deviation
\begin{equation}
  \langle\Delta H\%\rangle = \left\langle\frac{H_{\rm NM}-H_{\rm Markov}}{H_{\rm Markov}}\right\rangle\times100\%,
  \label{eq:meandev}
\end{equation}
where the average is taken over the redshifts of each dataset
(Table~\ref{tab:meandev}). The Power-law, Exponential and Gaussian kernels
raise $H(z)$ by $0.5$--$1.4\%$, consistent with $\sqrt{1+\Aeff}-1$. The largest
deviations, up to $2.7\%$, come from the Oscillatory kernel. The Linear kernel
departs from \LCDM{} by less than $0.9\%$ and by essentially zero for DR2.

\begin{table*}[!t]
  \caption{Mean percentage deviation $\langle\Delta H\%\rangle$,
  Eq.~\eqref{eq:meandev}, of each best-fit kernel from the Markovian baseline.}
  \label{tab:meandev}
  \begin{ruledtabular}
  \begin{tabular}{lrrrrr}
    Kernel & CC (\%) & DESI DR1 (\%) & DESI DR2 (\%) & CC+DR1 (\%) & CC+DR2 (\%) \\
    \hline
    Linear      & $+0.863$ & $+0.531$ & $-0.004$ & $+0.336$ & $+0.013$ \\
    Power law   & $+1.420$ & $+1.158$ & $+0.517$ & $+1.199$ & $+0.576$ \\
    Exponential & $+1.420$ & $+1.158$ & $+0.517$ & $+1.198$ & $+0.576$ \\
    Gaussian    & $+1.404$ & $+1.157$ & $+0.517$ & $+1.190$ & $+0.576$ \\
    Oscillatory & $+0.732$ & $+2.278$ & $+1.076$ & $+2.685$ & $+1.914$ \\
  \end{tabular}
  \end{ruledtabular}
\end{table*}

\section{Where the non-Markovian model performs best}
\label{sec:best}

Although none of the kernels is preferred over \LCDM{} across all datasets,
the degree to which the memory term improves the fit varies considerably
between kernels and datasets. In this section we identify the cases in which
the non-Markovian description performs best, and assess how far these cases
can be taken as support for cosmic memory.

The strongest performance is obtained for the Oscillatory kernel fitted to
DESI DR1. Here the memory term lowers the minimum $\chi^{2}$ from $8.40$ for
\LCDM{} to $0.33$, an improvement of $\dchi=8.07$. This is the only case in
which the non-Markovian model remains preferred after the additional
parameters are penalised, with $\Delta{\rm AIC}=-2.07$ and
$\Delta{\rm BIC}=-3.24$. The second-best case is the same kernel fitted to
DESI DR2, where $\chi^{2}_{\rm min}$ falls from $8.63$ to $1.01$
($\dchi=7.62$), giving a weak preference for memory with
$\Delta{\rm AIC}=-1.62$ and $\Delta{\rm BIC}=-2.24$. In both cases the
preferred oscillation frequency is similar, $\omega\simeq4$--$5$,
corresponding to a period of $\Delta z\simeq1.2$--$1.5$ in redshift lag.

These improvements must, however, be qualified. Both best fits require a
negative damping rate, $\alpha=-0.10$ for DR1 and $\alpha=-0.45$ for DR2,
which describes a memory kernel that grows rather than decays into the past.
Such behaviour is excluded by the microscopic derivation of
Sec.~\ref{sec:opensystem}, in which $\alpha$ is the width of a Lorentzian
spectral density and must be positive. The DR1 and DR2 samples also contain
only five and six points, so a three-parameter kernel can follow individual
data points closely; the value $\chi^{2}_{\rm min}=0.33$ for DR1 is itself an
indication of over-fitting. When the Cosmic Chronometer data are added, the
preference disappears: for CC+DR1 and CC+DR2 the Oscillatory kernel gives
$\Delta{\rm AIC}=+1.48$ and $+0.20$, and $\Delta{\rm BIC}=+6.15$ and $+4.95$,
both favouring \LCDM{}.

Among the kernels with a physically decaying memory, the best performance is
obtained for the Power-law, Exponential and Gaussian kernels fitted to the
combined CC+DR1 data, which improve the fit by $\dchi=1.48$. This gain is too
small to compensate for the two additional parameters, and \LCDM{} remains
preferred with $\Delta{\rm AIC}=+2.52$. As shown in Sec.~\ref{sec:fits}, the
improvement arises from the short-memory limit, in which these kernels act as
a uniform rescaling of $H(z)$ by about $1\%$, equivalent to a small increase
in $H_0$ rather than a genuine memory effect. The Linear kernel performs best
on the same combination, but its improvement is only $\dchi=0.57$, and its
amplitude remains consistent with zero.

In summary, the non-Markovian model performs best for the Oscillatory kernel
applied to DESI data alone, where it is weakly preferred over \LCDM{}. This
preference relies on an unphysical growing kernel and on very small samples,
and it does not survive the inclusion of the Cosmic Chronometer data. For all
kernels with physically allowed parameters, the memory term improves the fit
only marginally and is not required by current expansion-history data. The
recurring preference for $\omega\simeq4$--$5$ in both DESI releases is
nevertheless worth revisiting with the full DESI likelihood and with the
constraint $\alpha>0$ imposed, since a genuine oscillatory feature in $H(z)$
would be a distinctive signature of a resonant hidden sector.

\section{Discussion and conclusion}
\label{sec:conclusion}

We have explored a non-Markovian extension of Friedmann cosmology in which the
present expansion rate retains knowledge of its past through a causal memory
integral, Eq.~\eqref{eq:nm_friedmann}. Five continuous kernels (Linear, Power
law, Exponential, Gaussian and Oscillatory) were confronted with 30 Cosmic
Chronometer measurements and with DESI BAO data from DR1 and DR2, separately and
in combination, using multi-start $\chi^{2}$ minimisation, MCMC sampling and
information-criterion model comparison.

On the theoretical side, we showed that the five kernels are not independent
phenomenological choices. Each is the cosine transform,
Eq.~\eqref{eq:kernel_spectral}, of the spectral density of an unresolved
environment coupled to the expansion history, which endows the parameters
$(A,\alpha,\beta,\sigma,\omega)$ with interpretations as the coupling strength,
inverse correlation time, spectral index, bandwidth and resonance frequency of a
hidden sector. The Exponential kernel also follows from causal Israel--Stewart
bulk viscosity, with $\alpha=1/\tau$ the relaxation rate, while the Linear kernel
arises as the short-lag limit of the retarded Green's function of Deser--Woodard
non-local gravity. We further showed that in the short-memory limit all
decaying kernels reduce to a single integrated amplitude, $\Aeff$, which acts as
a rescaling of $H_0$ [Eq.~\eqref{eq:shortmemory}].

On the observational side, our main findings are as follows.
(i)~\LCDM{} with Planck 2018 parameters already fits all datasets well
($\chi^{2}=14.9$, $8.4$ and $8.6$ for CC, DR1 and DR2).
(ii)~The Linear amplitude is consistent with zero, with
$A=0.002\pm0.030$ for CC+DR2.
(iii)~The Power-law, Exponential and Gaussian kernels are observationally
degenerate. The data place them in the short-memory limit, where only
$\Aeff\simeq0.01$--$0.03$ is constrained, equivalent to a $0.5$--$1.5\%$ uniform
increase of $H(z)$ or a shift of $H_0$ by at most $\simeq1~\kms$ at fixed
background. They improve $\chi^{2}$ by at most $1.5$ and are disfavoured by the
information criteria. Their MCMC posteriors give only lower limits on the
decay rates ($\alpha,\beta\gtrsim50$) and exclude broad Gaussian memory. In the
Israel--Stewart picture this constrains the bulk viscosity $\zeta\propto\Aeff$
but leaves the relaxation time undetermined.
(iv)~The Oscillatory kernel improves the fit by $\dchi=4.5$--$8.1$ with a
frequency $\omega\simeq4$--$5$ common to the DESI datasets, but only with a
growing envelope ($\alpha<0$) that is inconsistent with its microscopic
derivation, and it is not preferred once CC data are included.
We conclude that current expansion-history data allow cosmic memory at the
percent level in $H(z)$ but do not require it. The memory term is therefore best
regarded as a physically motivated, testable extension of \LCDM{} that is
presently constrained rather than detected.

Several limitations should be noted. The background parameters were held at
their Planck 2018 values, which, together with the degeneracy of
Eq.~\eqref{eq:shortmemory}, means that part of the memory signal is
interchangeable with $H_0$. $H(z)$ was derived from $D_H/r_d$ alone, neglecting
its correlation with $D_M/r_d$ and the isotropic $D_V/r_d$ measurements. The
posterior extent of the degenerate kernels depends on the prior ranges, and the
Gaussian chains, at about $36\,\tau_{\rm int}$, would benefit from longer runs.

Natural extensions include: (i)~freeing $H_0$ and $\Omega_m$, which will
determine whether any memory signal survives once the $H_0$ degeneracy is
marginalised over; (ii)~imposing $\alpha>0$ for the Oscillatory kernel, as
required by Eq.~\eqref{eq:J_osc}, and sampling in $\Aeff$ with logarithmic
priors on the decay scales; (iii)~using the full DESI likelihood including
$D_M/r_d$, $D_V/r_d$ and their covariance, together with supernova data;
(iv)~computing the Bayesian evidence; (v)~a study of the effective equation of
state $w_{\rm eff}(z)$, Eq.~\eqref{eq:weff}, the deceleration parameter and
statefinder diagnostics; and (vi)~a perturbation-level formulation of the
memory term, since the growth of structure would provide an independent test
of the kernels. The full Deser--Woodard Green's function for the fiducial
$a(t)$ should also be computed numerically to determine whether the Power-law
and Gaussian forms can be obtained from non-local gravity.

\begin{acknowledgments}
PR acknowledges the Inter-University Centre for Astronomy and Astrophysics
(IUCAA), Pune, India, for the award of a visiting associateship.
\end{acknowledgments}



\begin{thebibliography}{99}

\bibitem{Zwanzig1961}
R.~Zwanzig, Phys.\ Rev.\ \textbf{124}, 983 (1961).

\bibitem{Mori1965}
H.~Mori, Prog.\ Theor.\ Phys.\ \textbf{33}, 423 (1965).

\bibitem{Caldeira1983a}
A.~O.~Caldeira and A.~J.~Leggett, Physica A \textbf{121}, 587 (1983).

\bibitem{Deser2007}
S.~Deser and R.~P.~Woodard, Phys.\ Rev.\ Lett.\ \textbf{99}, 111301 (2007),
arXiv:0706.2151.

\bibitem{Israel1976}
W.~Israel, Ann.\ Phys.\ (N.Y.) \textbf{100}, 310 (1976).

\bibitem{IsraelStewart1979}
W.~Israel and J.~M.~Stewart, Ann.\ Phys.\ (N.Y.) \textbf{118}, 341 (1979).

\bibitem{Weinberg2008}
S.~Weinberg, \textit{Cosmology} (Oxford University Press, Oxford, 2008).

\bibitem{Capozziello2011}
S.~Capozziello and M.~De~Laurentis, Phys.\ Rep.\ \textbf{509}, 167 (2011).

\bibitem{DESI2024}
A.~G.~Adame \textit{et al.} (DESI Collaboration), \textit{DESI 2024 VI:
Cosmological constraints from the measurements of baryon acoustic
oscillations}, J.\ Cosmol.\ Astropart.\ Phys.\ \textbf{02} (2025) 021,
arXiv:2404.03002.

\bibitem{DESI2025}
M.~Abdul~Karim \textit{et al.} (DESI Collaboration), \textit{DESI DR2 results.
II. Measurements of baryon acoustic oscillations and cosmological constraints},
Phys.\ Rev.\ D \textbf{112}, 083515 (2025), arXiv:2503.14738.

\bibitem{Planck2018}
N.~Aghanim \textit{et al.} (Planck Collaboration), Astron.\ Astrophys.\
\textbf{641}, A6 (2020).

\bibitem{Caldeira1983b}
A.~O.~Caldeira and A.~J.~Leggett, Ann.\ Phys.\ (N.Y.) \textbf{149}, 374 (1983).

\bibitem{Weiss2012}
U.~Weiss, \textit{Quantum Dissipative Systems}, 4th ed.\ (World Scientific,
Singapore, 2012).

\bibitem{Kubo1966}
R.~Kubo, Rep.\ Prog.\ Phys.\ \textbf{29}, 255 (1966).

\bibitem{Hiscock1983}
W.~A.~Hiscock and L.~Lindblom, Ann.\ Phys.\ (N.Y.) \textbf{151}, 466 (1983).

\bibitem{Hiscock1985}
W.~A.~Hiscock and L.~Lindblom, Phys.\ Rev.\ D \textbf{31}, 725 (1985).

\bibitem{Deffayet2009}
C.~Deffayet and R.~P.~Woodard, J.\ Cosmol.\ Astropart.\ Phys.\ \textbf{08}
(2009) 023, arXiv:0904.0961.

\bibitem{Nojiri2008}
S.~Nojiri and S.~D.~Odintsov, Phys.\ Lett.\ B \textbf{659}, 821 (2008),
arXiv:0708.0924.

\bibitem{Maggiore2014}
M.~Maggiore and M.~Mancarella, Phys.\ Rev.\ D \textbf{90}, 023005 (2014),
arXiv:1402.0448.

\bibitem{Dirian2015}
Y.~Dirian, S.~Foffa, M.~Kunz, M.~Maggiore, and V.~Pettorino,
J.\ Cosmol.\ Astropart.\ Phys.\ \textbf{04} (2015) 044, arXiv:1411.7692.

\bibitem{Dirian2016}
Y.~Dirian, S.~Foffa, M.~Kunz, M.~Maggiore, and V.~Pettorino,
J.\ Cosmol.\ Astropart.\ Phys.\ \textbf{05} (2016) 068, arXiv:1602.03558.

\bibitem{Simon2005}
J.~Simon, L.~Verde, and R.~Jimenez, Phys.\ Rev.\ D \textbf{71}, 123001 (2005),
arXiv:astro-ph/0412269.

\bibitem{Stern2010}
D.~Stern, R.~Jimenez, L.~Verde, M.~Kamionkowski, and S.~A.~Stanford,
J.\ Cosmol.\ Astropart.\ Phys.\ \textbf{02} (2010) 008, arXiv:0907.3149.

\bibitem{Moresco2012a}
M.~Moresco \textit{et al.}, J.\ Cosmol.\ Astropart.\ Phys.\ \textbf{08} (2012)
006, arXiv:1201.3609.

\bibitem{Zhang2014}
C.~Zhang, H.~Zhang, S.~Yuan, S.~Liu, T.-J.~Zhang, and Y.-C.~Sun,
Res.\ Astron.\ Astrophys.\ \textbf{14}, 1221 (2014), arXiv:1207.4541.

\bibitem{Moresco2015}
M.~Moresco, Mon.\ Not.\ R.\ Astron.\ Soc.\ \textbf{450}, L16 (2015),
arXiv:1503.01116.

\bibitem{Moresco2016}
M.~Moresco \textit{et al.}, J.\ Cosmol.\ Astropart.\ Phys.\ \textbf{05} (2016)
014, arXiv:1601.01701.

\bibitem{Ratsimbazafy2017}
A.~L.~Ratsimbazafy, S.~I.~Loubser, S.~M.~Crawford, C.~M.~Cress, B.~A.~Bassett,
R.~C.~Nichol, and P.~V\"ais\"anen, Mon.\ Not.\ R.\ Astron.\ Soc.\
\textbf{467}, 3239 (2017), arXiv:1702.00418.

\bibitem{Jimenez2002}
R.~Jimenez and A.~Loeb, Astrophys.\ J.\ \textbf{573}, 37 (2002),
arXiv:astro-ph/0106145.

\bibitem{Moresco2012b}
M.~Moresco, L.~Verde, L.~Pozzetti, R.~Jimenez, and A.~Cimatti,
J.\ Cosmol.\ Astropart.\ Phys.\ \textbf{07} (2012) 053, arXiv:1201.6658.

\bibitem{DESI2024III}
A.~G.~Adame \textit{et al.} (DESI Collaboration), \textit{DESI 2024 III: Baryon
acoustic oscillations from galaxies and quasars}, J.\ Cosmol.\ Astropart.\
Phys.\ \textbf{04} (2025) 012, arXiv:2404.03000.

\bibitem{DESI2024IV}
A.~G.~Adame \textit{et al.} (DESI Collaboration), \textit{DESI 2024 IV: Baryon
acoustic oscillations from the Lyman alpha forest}, J.\ Cosmol.\ Astropart.\
Phys.\ \textbf{01} (2025) 124, arXiv:2404.03001.

\bibitem{DESI2025Lya}
M.~Abdul-Karim \textit{et al.} (DESI Collaboration), \textit{DESI DR2 results.
I. Baryon acoustic oscillations from the Lyman alpha forest},
Phys.\ Rev.\ D \textbf{112}, 083514 (2025), arXiv:2503.14739.

\bibitem{emcee}
D.~Foreman-Mackey, D.~W.~Hogg, D.~Lang, and J.~Goodman, Publ.\ Astron.\ Soc.\
Pac.\ \textbf{125}, 306 (2013), arXiv:1202.3665.

\bibitem{getdist}
A.~Lewis, arXiv:1910.13970 (2019).

\end{thebibliography}
\end{document}